\documentclass[prl,twocolumn,nofootinbib,longbibliography]{revtex4-2}
\usepackage{amsmath,amssymb}
\usepackage[dvipsnames]{xcolor}
\usepackage{hyperref}
\usepackage{tikz}
\hypersetup{
  colorlinks=true,
  linkcolor=MidnightBlue,
  citecolor=MidnightBlue,
  urlcolor=MidnightBlue
}

\usepackage{bm}
\usepackage{mathrsfs}

\newcommand{\Z}{\mathbb{Z}}
\newcommand{\SU}{\mathrm{SU}}

\newcommand{\cH}{\mathcal{H}}
\newcommand{\Zgrav}{Z_{\text{grav}}}
\newcommand{\Ztop}{Z_{\text{top}}}
\newcommand{\Ztot}{Z_{\text{tot}}}

\newcommand{\DD}{\mathrm{D}}
\newcommand{\ii}{\mathrm{i}}
\newcommand{\ee}{\mathrm{e}}
\newcommand{\rev}[1]{{\color{black}#1}}

\begin{document}

\title{When do real observers resolve de Sitter's imaginary problem?}

\author{Ahmed Farag Ali}
\affiliation{Essex County College, Newark, NJ 07102, USA}
\affiliation{Department of Physics, Benha University, 13511 Benha, Egypt}

\date{\today}

\begin{abstract}
The universal phase $\rev{\ii}^{D+2}$ of the Euclidean de Sitter path integral obstructs a straightforward state-counting interpretation of the Gibbons--Hawking entropy. Building on Maldacena's proposal that specific black-hole observers can reorganize this phase, we derive a general constraint on when such ``real observers'' can succeed. By distinguishing \emph{gravitational observers} from \emph{topological spectators}, we show at quadratic semiclassical order that any sector whose \emph{infrared effective} action is metric independent at the de Sitter saddle factorizes in the path integral, $\Ztot = \Zgrav^{(\text{obs})}\Ztop$, so the imaginary phase persists regardless of the sector's information-processing capabilities. Using confining $\SU(3)$ gauge theory and topological orders as examples, we demonstrate that an information-bearing clock is necessary but insufficient: only observers whose fluctuations share the negative modes of the conformal factor belong to the special class that can remove the de Sitter phase.
\end{abstract}

\maketitle

\paragraph{Motivation.---}
The Euclidean path integral on the round de Sitter sphere $S^D$,
\begin{equation}
  \Zgrav(S^D)
  =
  \int \rev{\DD}g \, \rev{\ee}^{-I_{\text{EH}}[g]},
\end{equation}
has long been known to carry subtle pathologies. While Gibbons and Hawking showed that the saddle reproduces the thermodynamic entropy of the cosmological horizon,\cite{GibbonsHawking1977} Gibbons, Hawking and Perry noted the indefinite sign of the gravitational action,\cite{GibbonsHawkingPerry1978} and Polchinski explicitly identified $D+2$ negative modes in the conformal sector which generate a universal phase:\cite{Polchinski1989}
\begin{equation}
  \Zgrav(S^D)
  \;\sim\;
  \exp\!\left(\frac{A_c}{4G_N}\right)\,
  \rev{\ii}^{D+2}\,
  \big|\det\nolimits'(\Delta_{\text{grav}})\big|^{-1/2}.
  \label{eq:polchinski}
\end{equation}
Here $A_c$ is the area of the cosmological horizon. The sign is such that $I_{\text{EH}}[g_{\text{dS}}] = - A_c/(4G_N)$, so that $\rev{\ee}^{-I_{\text{EH}}[g_{\text{dS}}]}$ has the usual entropy-counting form.

To resolve the interpretational difficulties posed by the phase $\rev{\ii}^{D+2}$, Maldacena recently proposed a concrete remedy: by coupling gravity to a charged near-extremal black hole equipped with a physical clock, the additional negative modes of the observer reorganize the path integral so that the static-patch density of states becomes real.\cite{Maldacena2024RealObservers}

This proposal has catalyzed a wave of precise investigations into the nature of observers and phases in de Sitter gravity. Regarding the phase structure, Ivo, Maldacena and Sun traced the Euclidean phase to specific physical instabilities of the conformal factor,\cite{IvoMaldacenaSun2025} while Shi and Turiaci generalized this analysis to arbitrary Einstein manifolds.\cite{ShiTuriaci2025PhaseGPI} Chen, Stanford, Tang and Yang emphasized a sharp tension, showing that the Euclidean path integral carries a robust complex phase that resists simple cancellation, though they argue that a positive density of states is nevertheless recoverable,\cite{ChenEtAl2025PhaseDensity} building on earlier classifications of sphere path integrals by Law\cite{Law2021SphereInts} and explicit entropy computations by Anninos \emph{et al.}\cite{AnninosEtAl2022QdS}

Simultaneously, the definition of the observer itself has been formalized algebraically. Harlow, Usatyuk and Zhao identified observers with subalgebras and cyclic states in a closed-universe Hilbert space;\cite{HarlowUsatyukZhao2025} Chen further developed this by modeling observers as ``sources'' for gravitational Hilbert spaces;\cite{Chen2025ObserversHilbert} and Chen and Xu constructed an explicit algebra for covariant observers.\cite{ChenXu2025CovariantObservers} In the holographic context, Akers \emph{et al.} analyzed how such observers map to dual descriptions,\cite{AkersEtAl2025HolographicMaps} while Chandrasekaran, Longo, Penington and Witten provided an algebra of observables adapted to the horizon, distinguishing static observers from the global geometry.\cite{ChandrasekaranLongoPeningtonWitten2023} Related flow-geometry developments include the centaur-algebra of observables for asymptotically AdS spacetimes with infalling-observer protocols, and a two-dimensional flow-geometry microstate construction reproducing the Gibbons--Hawking entropy.\cite{Aguilar-Gutierrez:2023odp,Espindola:2025wjf}. From a constraints perspective, Horowitz, Marolf and Santos proved that the gravitational constraints alone are insufficient to fix the quantum theory, necessitating explicit observer data.\cite{HorowitzMarolfSantos2025Constraints} Finally, microscopic proposals in which $\SU(3)$ confinement fixes the vacuum energy and nature's constants offer a concrete realization of gravitational clocks tied to the fundamental vacuum architecture.\cite{Ali:2024rnw,Ali:2025wld}. For an $\SU(3)$ clock realization in de Sitter, see Ref.~\onlinecite{Ali:2026vwk}.

These developments suggest a natural conceptual picture: once an observer with a clock is included, the de Sitter phase might be brought under control. The main purpose of this Letter is to make this picture precise by identifying which ingredients---gravitational backreaction, informational clocks, or topological structure---are actually responsible for changing the phase, and to formulate this as a criterion on which observers qualify as genuinely gravitational.

\paragraph{Three notions.---}
We distinguish three logically independent notions that frequently appear under the single label ``observer''.

\emph{Worldlines.}  
Any localized subsystem defines a worldline (or worldvolume) in spacetime: examples range from cosmic strings and monopoles to anyons in a quantum Hall fluid or vortices in a superconductor.\cite{VilenkinShellard1994,WenNiu1990QHTopDegeneracy,Zurek1985CosmoSuperfluidHelium} A worldline is a geometric object; by itself it does not yet encode any temporal ordering of internal states.

\emph{Informational clocks.}  
In relational and informational approaches to time, a ``clock'' is a subsystem whose internal state provides an intrinsic ordering of distinguishable events.\cite{PageWootters1983,ConnesRovelli1994} Concretely, we say a subsystem along a worldline carries an informational clock if it has an internal Hilbert space $\cH_{\text{clock}}$ and undergoes a sequence of operations
\begin{equation}
  U_n : \cH_{\text{clock}} \to \cH_{\text{clock}},
\end{equation}
such that the states $\{U_n|\psi_0\rangle\}_{n\in\mathbb{Z}}$ are distinguishable and naturally ordered, with $n$ serving as a logical or circuit-time parameter. This framework encompasses Page--Wootters relational time, thermal-time constructions, and circuit-based notions of time.\cite{PageWootters1983,ConnesRovelli1994}. Many topological and condensed-matter sectors supply such clocks very naturally: braiding anyons in a fractional quantum Hall state,\cite{WenNiu1990QHTopDegeneracy} acting with stabilizers in a toric code,\cite{Kitaev2003ToricCode} or tracking reconnection events in a defect network.\cite{VilenkinShellard1994,Zurek1985CosmoSuperfluidHelium}

\emph{Gravitational observers.}  
The Euclidean de Sitter phase problem, however, is a statement about the index of the Hessian of the gravitational action.\cite{Polchinski1989,ChenEtAl2025PhaseDensity,IvoMaldacenaSun2025,ShiTuriaci2025PhaseGPI} We call a subsystem a \emph{gravitational observer} if its action $I_{\text{obs}}[g,\gamma,\Phi]$ has a localized stress tensor that is non-negligible at the de Sitter scale and whose quadratic couplings to metric fluctuations $h_{\mu\nu}$ mix with the conformal sector. In other words, the combined Hessian
\begin{equation}
  \Delta_{\text{grav+obs}}
\end{equation}
has a different index (number of negative directions) from $\Delta_{\text{grav}}$ alone. Maldacena's black-hole observer is an explicit and well-controlled realization: the additional negative modes live in a joint configuration space of metric and worldline degrees of freedom, and their net effect is to alter the overall phase in a carefully constructed static-patch partition function\cite{Maldacena2024RealObservers}.

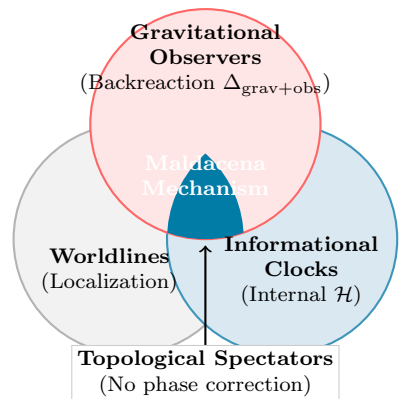
\begin{figure}[t]
\centering
\begin{tikzpicture}[scale=0.85, every node/.style={align=center, font=\footnotesize}]

  \def\R{1.8} 

  \coordinate (C1) at (0,0);       
  \coordinate (C2) at (2.4,0);     
  \coordinate (C3) at (1.2,1.8);   

  \fill[gray!10] (C1) circle (\R);
  \draw[gray!60, thick] (C1) circle (\R);
  
  \fill[MidnightBlue!10] (C2) circle (\R);
  \draw[MidnightBlue!60, thick] (C2) circle (\R);
  
  \fill[red!10] (C3) circle (\R);
  \draw[red!60, thick] (C3) circle (\R);

  \begin{scope}
    \clip (C1) circle (\R);
    \clip (C2) circle (\R);
    \fill[MidnightBlue!80] (C3) circle (\R);
  \end{scope}

  \node at (-0.3,-0.5) {\textbf{Worldlines}\\(Localization)};
  \node at (2.7,-0.5) {\textbf{Informational}\\\textbf{Clocks}\\(Internal $\mathcal{H}$)};
  \node at (1.2,2.8) {\textbf{Gravitational}\\\textbf{Observers}\\(Backreaction $\Delta_{\text{grav+obs}}$)};

  \node[text=white] at (1.2, 1.0) {\textbf{Maldacena}\\\textbf{Mechanism}};

  \draw[->, thick, black] (1.2, -1.8) -- (1.2, -0.1);
  \node[fill=white, inner sep=2pt, draw=black!20] at (1.2, -2.1) 
    {\textbf{Topological Spectators}\\(No phase correction)};

\end{tikzpicture}
\caption{
Schematic relation between worldlines, informational clocks, and gravitational observers. The Maldacena mechanism operates only in the triple intersection, where a system both carries a clock and backreacts on the conformal factor so as to share its negative modes. Sectors that have worldlines and clocks but decouple from the metric (bottom overlap) are topological spectators: they can store and order information, but they cannot modify the universal de Sitter phase.
}
\label{fig:venn}
\end{figure}

These three notions are independent: a system may have worldlines without a clock; a clock without appreciable gravitational coupling; or strong gravitational coupling without a large internal Hilbert space. The mechanism of Ref.~\onlinecite{Maldacena2024RealObservers} implicitly uses the full intersection: a worldline carrying an informational clock whose fluctuations also share the conformal negative modes, as illustrated schematically in Fig.~\ref{fig:venn}.

\paragraph{Topological spectators and factorization.---}
We now identify a broad class of sectors that, while often rich from the informational point of view, do not affect the de Sitter phase at the level of semiclassical path integrals.

By a \emph{topological spectator} we mean a sector with fields $A$ whose stress-energy response functions vanish at macroscopic scales. While the microscopic (bare) action $I_{\text{top}}[A;g]$ may depend on the metric, we require that, after integrating out gapped excitations and passing to the infrared effective action $I_{\text{top}}^{(\text{IR})}[A;g]$ at scales $\sim R_{\text{dS}}$, the induced stress-tensor fluctuations decouple from the geometry. At the quadratic level, this decoupling can be summarized schematically as
\begin{equation}
  \frac{\delta T_{\mu\nu}}{\delta g_{\rho\sigma}}
  \;\propto\;
  \left.
  \frac{\delta^2 I_{\text{top}}^{(\text{IR})}}{\delta g_{\mu\nu}\,\delta g_{\rho\sigma}}
  \right|_{g_{\text{dS}}}
  \simeq 0,
\end{equation}
so that metric fluctuations and spectator fluctuations do not mix in the total Hessian.

These conditions can be phrased as the following working assumption:

\emph{Hypothesis (spectator decoupling).}  
Let $I_{\text{top}}^{(\text{IR})}[A;g]$ be a diffeomorphism-invariant infrared effective functional whose explicit dependence on $g_{\mu\nu}$ at the de Sitter saddle is purely topological, in the sense that $\delta^2 I_{\text{top}}^{(\text{IR})}/\delta g\,\delta g|_{g_{\text{dS}}}=0$ up to corrections suppressed by powers of $\ell/R_{\text{dS}}$. Then, after gauge fixing the metric and including the associated Faddeev--Popov ghosts, the full gauge-fixed quadratic form around $g_{\text{dS}}$ is block-diagonal between $(h_{\mu\nu},\text{ghosts},\text{obs})$ and the spectator sector $A$.

Including both an observer sector and a topological spectator, the Euclidean path integral on $S^D$ takes the schematic form
\begin{equation}
  \Ztot(S^D)
  =
  \int\!\rev{\DD}g\, \rev{\ee}^{-I_{\text{EH}}[g]} \,
  Z_{\text{obs}}[g] \,
  Z_{\text{top}}[g],
  \label{eq:Z_total}
\end{equation}
where
\begin{equation}
\begin{split}
  Z_{\text{obs}}[g]
  &=
  \int\!\rev{\DD}\gamma\,\rev{\DD}\Phi\, \rev{\ee}^{-I_{\text{obs}}[g,\gamma,\Phi]},
  \\
  Z_{\text{top}}[g]
  &=
  \int\!\rev{\DD}A\, \rev{\ee}^{-I_{\text{top}}^{(\text{IR})}[A;g,\dots]}.
\end{split}
\label{eq:Z_components}
\end{equation}
Under the spectator hypothesis, $Z_{\text{top}}[g]$ is insensitive to small variations of $g$ about $g_{\text{dS}}$ and can be replaced, within the saddle-point expansion, by the constant $Z_{\text{top}}(S^D) = Z_{\text{top}}[g_{\text{dS}}]$. 

Expanding around $g_{\text{dS}}$ and integrating over Gaussian fluctuations, one finds that the Hessian is block-diagonal,
\begin{equation}
  \Delta_{\text{tot}}
  =
  \begin{pmatrix}
    \Delta_{\text{grav+obs}} & 0 \\
    0 & \Delta_{\text{top}}
  \end{pmatrix},
\end{equation}
and hence the partition function factorizes:
\begin{equation}
  \Ztot(S^D)
  =
  \Zgrav^{(\text{obs})}(S^D)\,\Ztop(S^D),
  \label{eq:factorization}
\end{equation}
with
\begin{equation}
\begin{aligned}
  \Zgrav^{(\text{obs})}(S^D)
    &= \rev{\ee}^{-I_{\text{eff}}[g_{\star}]}\,
       \det\nolimits'^{-1/2}(\Delta_{\text{grav+obs}}),
  \\
  \Ztop(S^D)
    &= \det\nolimits^{-1/2}(\Delta_{\text{top}}).
\end{aligned}
\end{equation}
Here $g_{\star}$ denotes the relevant saddle including the observer sector. Writing $\Ztop = |\Ztop| \rev{\ee}^{\rev{\ii}\varphi_{\text{top}}}$, we see that the spectator contributes a fixed phase $\varphi_{\text{top}}$ determined by topological data, but does not change the index of the conformal sector. If no gravitational observer is present, $\Zgrav^{(\text{obs})}$ reduces to \eqref{eq:polchinski} with phase $\rev{\ii}^{D+2}$. If a gravitational observer is present, its effect on the phase is entirely contained in $\Zgrav^{(\text{obs})}$; the spectator only multiplies the result by a constant complex factor.

Corrections that violate the hypothesis are naturally organized in powers of $\varepsilon \sim (\ell/R_{\text{dS}})^p$ for some positive $p$: they can shift eigenvalues slightly but do not change the number of negative modes until $\varepsilon$ becomes order unity, at which point the sector ceases to be a spectator in any case. This is the precise sense in which such sectors are spectators with respect to the de Sitter phase: they may carry nontrivial Hilbert spaces and robust informational clocks, but they do not participate in the reorganization of negative modes that underlies the mechanism of Ref.~\onlinecite{Maldacena2024RealObservers}.

\paragraph{$\bm{\SU(3)}$ as spectator and as clock.---}
$\SU(3)$ Yang--Mills theory offers a concrete illustration of the distinction between informational and gravitational sectors. The Euclidean action on $S^4$ can be written in differential-form notation as
\begin{equation}
  I_{\text{YM}}
  =
  \int_{S^4} \left[ \frac{1}{2g_{\text{YM}}^2} \,\text{Tr}(F \wedge \star F)
  +
  \frac{i\theta}{8\pi^2} \,\text{Tr}(F\wedge F) \right].
\end{equation}
The topological $\theta$-term is strictly metric independent, as the wedge product relies only on the differentiable structure of the manifold; its variation with respect to $g_{\mu\nu}$ vanishes identically. The kinetic term depends on the metric via the Hodge star, but in the confining phase the spectrum develops a mass gap $m_{\text{gap}} \sim \ell_{\text{YM}}^{-1}$. Provided $R_{\text{dS}} \gg \ell_{\text{YM}}$, stress-tensor fluctuations $\langle \delta T_{\mu\nu}(x) \delta T_{\rho\sigma}(y) \rangle$ are exponentially suppressed by factors of $\rev{\ee}^{-m_{\text{gap}} d(x,y)}$ on de Sitter scales. Consequently, the metric-dependent excitations can be integrated out, leaving only vacuum-energy renormalization and an effectively topological infrared action $I_{\text{top}}^{(\text{IR})}$.

In this regime the path integral factorizes into the gravitational sector and a discrete sum over instanton numbers $k \in \pi_3(\SU(3)) \cong \Z$:
\begin{equation}
  \Ztop(S^4) \approx \sum_{k\in\Z} n_k \, \rev{\ee}^{-S_k} \rev{\ee}^{\rev{\ii}\theta k}.
\end{equation}
The resulting phase is determined purely by $\theta$ and the measure degeneracies $n_k$, and is algebraically independent of the conformal factor's unstable modes.

Equivalently, this physics can be captured by thin $\SU(3)$ center vortices carrying $\Z_3$ magnetic flux. When their tension is negligible relative to the de Sitter scale, these vortices act as test objects: their contributions to the path integral are topological invariants, such as linking and intersection numbers. These worldvolumes support a finite-dimensional Hilbert space of $\Z_3$ charge sectors, allowing them to function as informational clocks or logical qubits. However, because their backreaction on the metric is suppressed at the de Sitter radius, they remain topological spectators and cannot resolve the factor $\rev{\ii}^{D+2}$ in Eq.~\eqref{eq:polchinski}.

By contrast, the situation changes in microscopic models where $\SU(3)$ confinement defines the vacuum architecture and fixes the cosmological constant and nature's constants\cite{Ali:2024rnw,Ali:2025wld}. In this regime, the gauge degrees of freedom are no longer spectators living on a background; they constitute part of the background geometry. The fluctuations of the confining condensate mix directly with the metric sector and contribute to $\Delta_{\text{grav+obs}}$, promoting the $\SU(3)$ sector from a topological spectator to a true gravitational observer capable of modifying the unstable contour and the associated phase.

\paragraph{Spectator universality class.---}
The structure leading to \eqref{eq:factorization} is not special to $\SU(3)$ and appears in many familiar systems. Topological field theories such as Chern--Simons and BF theories provide effective descriptions of quantum Hall states, topological insulators and other topologically ordered phases.\cite{WenNiu1990QHTopDegeneracy,QiHughesZhang2008TIFieldTheory} Their bulk actions are topological functionals of gauge fields, independent of local metric fluctuations, and their nontrivial physics resides in anyon braiding and boundary modes. Lattice topological orders and quantum memories, such as toric-code models, reduce at long distances to similar topological gauge theories.\cite{Kitaev2003ToricCode} Dilute networks of cosmic strings and other defects, treated in a strict decoupling limit, likewise contribute only topological phases and selection rules to the Euclidean path integral.\cite{VilenkinShellard1994,Zurek1985CosmoSuperfluidHelium}

These examples provide a broad and physically motivated universality class of sectors that satisfy the spectator-decoupling hypothesis and realize the factorized structure \eqref{eq:factorization} in concrete systems.

\paragraph{Scope and Discussion.---}
The distinction between gravitational observers and topological spectators places a sharp physical bound on where the Maldacena mechanism can operate. The near-extremal black-hole observer of Ref.~\onlinecite{Maldacena2024RealObservers} succeeds because its mass and compactness allow its stress tensor to modify the de Sitter saddle at order unity. By contrast, most of the matter content of a $\Lambda$-dominated universe---cold dark matter perturbations, baryonic inhomogeneities, or dilute defects---behave as informational clocks yet weakly backreacting spectators. To leading semiclassical order, they contribute a spectator factor $\Ztop$ that leaves the universal gravitational phase $\rev{\ii}^{D+2}$ intact.

This constraint extends naturally to algebraic and holographic perspectives on gravity in a closed universe.\cite{HarlowUsatyukZhao2025,Chen2025ObserversHilbert,ChenXu2025CovariantObservers,ChandrasekaranLongoPeningtonWitten2023,AkersEtAl2025HolographicMaps}. Enlarging an observer algebra by adjoining internal symmetry sectors, topological orders and code subspaces is essential for information-theoretic questions, yet it leaves the Euclidean de Sitter phase unchanged so long as these sectors do not couple to the conformal factor at quadratic order. The mechanism is genuinely gravitational: it singles out clocks whose fluctuations share the same unstable directions as the metric. Consequently, only sufficiently compact astrophysical objects, or microscopic degrees of freedom appearing in proposals that fix the vacuum energy and nature's constants (as in the confining scenarios of Refs.~\onlinecite{Ali:2024rnw,Ali:2025wld}), belong to the special class capable of resolving the imaginary problem.

Conceptually, this disentangles three statements: (i) there exists a worldline; (ii) there exists an internal, informational clock; (iii) there exists a gravitational clock that changes the de Sitter phase. Our contribution is to specify the conditions under which (i) and (ii) are insufficient. We conclude that the act of observation in quantum gravity is not merely the recording of information, but a dynamical process of gravitational backreaction. An observer who does not disturb the geometry cannot by itself resolve the Euclidean de Sitter phase problem. It would be interesting to repeat this analysis in lower-dimensional models, such as de Sitter analogues of Jackiw--Teitelboim gravity coupled to topological matter, and to explore scenarios where topological sectors flow from being spectators to defining the background geometry itself. We hope to report on these in the future.

\bibliographystyle{apsrev4-1}
\bibliography{refs}

\end{document}